\documentclass[twocolumn]{aas63_mod}
\usepackage[caption=false]{subfig}
\usepackage{booktabs}
\usepackage[T1]{fontenc}
\usepackage{lineno}
\newcommand{\teff}{\ensuremath{T_{\mathrm{eff}}}}
\newcommand{\logg}{\ensuremath{\log g}}
\newcommand{\logy}{\ensuremath{\log y}}
\definecolor{rotundaorange}{rgb}{0.898,0.44705882352,0}
\definecolor{pantherpurple}{rgb}{0.2,0,0.447058}
\definecolor{remeisblue}{rgb}{0, 0, 0.5450980}
\definecolor{potsdamblue}{rgb}{0.06274509803,0.28235294117,0.4431372549}
\definecolor{chicagomaroon}{rgb}{0.52549,0.12156863,0.25490196}

\hypersetup{linkcolor=red,citecolor=blue}

\newcommand\target{\textit{Gaia} DR2 6097540197980557440}

\begin{document}

\title{Eclipsing Binaries Found by the EREBOS Project: \\ \textit{Gaia} DR2 6097540197980557440 -- A Deeply Eclipsing sdB+dM System}

\correspondingauthor{Kyle A. Corcoran}
\email{kac8aj@virginia.edu}

\author[0000-0002-2764-7248]{Kyle A. Corcoran}
\affiliation{University of Virginia, Department of Astronomy, 530 McCormick Rd., Charlottesville, VA 22904, USA}
\author[0000-0002-8558-4353]{Brad N. Barlow}
\affiliation{High Point University, Department of Physics, One University Parkway, High Point, NC 27268, USA}
\author[0000-0001-6339-6768]{Veronika Schaffenroth}
\affiliation{
Universit{\"a}t Potsdam, Institut f{\"u}r Physik und Astronomie, Haus 28, Karl-Liebknecht-Str. 24/25, 14476, Potsdam-Golm, Germany}
\author[0000-0001-7798-6769]{Uli Heber}
\affiliation{Dr.~Karl Remeis-Observatory \& ECAP, Astronomical Institute, Friedrich-Alexander University Erlangen-Nuremberg (FAU),
Sternwartstr.~7, 96049 Bamberg, Germany}
\author[0000-0001-8444-2305]{Stephen Walser}
\affiliation{Virginia Tech, Department of Physics, 850 West Campus Drive, Blacksburg, VA 24061, USA}
\author[0000-0002-0465-3725]{Andreas Irgang}
\affiliation{Dr.~Karl Remeis-Observatory \& ECAP, Astronomical Institute, Friedrich-Alexander University Erlangen-Nuremberg (FAU),
Sternwartstr.~7, 96049 Bamberg, Germany}






\begin{abstract}

We present time-series spectroscopy and photometry of \textit{Gaia} DR2 6097540197980557440, a new deeply--eclipsing hot subdwarf B (sdB) + M dwarf (dM) binary. We discovered this object during the course of the Eclipsing Reflection Effect Binaries from Optical Surveys (EREBOS) project, which aims to find new eclipsing sdB+dM binaries (HW Vir systems) and increase the small sample of studied systems.
In addition to the primary eclipse, which is in excess of $\sim$5 magnitudes in the optical, the light curve also shows features typical for other HW Vir binaries such as a secondary eclipse and strong reflection effect from the 
irradiated, cool companion.
The orbital period is 0.127037 d ($\sim$3 hr), falling right at the peak of the orbital period distribution of known HW Vir systems. Analysis of our time--series spectroscopy yields a radial velocity semi-amplitude of $K_{\rm sdB}=100.0\pm2.0\,{\rm km\,s}^{-1}$, which is amongst the fastest line--of--sight velocities found to date for an HW Vir binary. 
State-of-the-art atmospheric models that account for deviations from local thermodynamic equilibrium are used to determine the atmospheric parameters of the sdB.
Although we cannot claim a unique light curve modeling solution, the best--fitting model has an sdB mass of $M_{\rm sdB} = 0.47\pm0.03\,M_{\odot}$ and a companion mass of $M_{\rm dM} = 0.18\pm0.01\,M_{\odot}$. 
The radius of the companion appears to be inflated relative to theoretical mass--radius relationships, consistent with other known HW Vir binaries. Additionally, the M dwarf is one of the most massive found to date amongst this type of binary. 
\end{abstract}




\section{\textbf{Introduction}} \label{sec:intro}

Most hot subdwarfs are core He--burning extreme horizontal branch (EHB) stars that formed from red giant progenitors that experienced mass loss near the tip of the giant branch, due to binary interactions \citep[see][for a detailed review]{Heber2016}. With temperatures from $20,000-45,000\rm\,K$ and spectra dominated by broad H Balmer lines, they are classified either as sdB stars or sdOB stars if they display the He\,{\sc ii}
4686 \AA\ line. They show a tight mass distribution peaking near 0.47$M_{\odot}$ (the `canonical' mass) and have radii around 0.2$R_{\odot}$.  Theoretical models such as those in \citet{Han2002,Han2003} describe sdB formation scenarios that account for the mass loss in these systems, with three possible formation channels depending on the initial configuration and mass ratio of the binary.  One formation channel produces an sdB via Roche lobe overflow (RLOF) 
to a MS companion of K type and earlier. The binaries that form in this way are typically wide binaries ($P = 10-1500 \rm\, d$).  These systems are often called ``composite" binaries, 
as both stars are seen in the spectrum and account for 30 -- 40\% of all sdBs \citep[for an overview see,][]{Vos2019}. 
The rest of the sdBs do not show any signs of a companion in their spectra. \citet{Maxted2001} showed that a high fraction of those sdBs do exist in short-period binaries leading to radial velocity variations. Those can be formed by common envelope (CE) evolution, which produces close binaries with periods as short as $\sim$1.5 hr with a hot subdwarf primary and a cool, low-mass companion.  In this scenario, an evolving red giant and a companion object enter a CE, and the angular momentum resonant in the orbit of the binary is deposited into the envelope, ejecting it from the system.  Typically, this companion is stellar in nature; however, \citet{Soker1998} proposed that sub-stellar and even planetary mass companions could be sufficient to provide the orbital angular momentum necessary to eject the envelope \citep[e.g.,][]{schaffenroth2015}. The remaining sdBs do not show any radial velocity variations and appear hence single. Such single sdBs could be formed by the merger of two He WDs. Another possibility is that a substellar companion was responsible for the mass loss, which was destroyed during the common envelope phase.

The main challenge in studying close sdB binaries and their properties comes from the single-lined nature of these systems, allowing only for mass limits inferred based on the proposed inclination; however, some systems benefit from the presence of an eclipse, which helps to constrain the inclination and allows for more precise mass measurements.  These so called HW Vir systems also show photometric variation due to the reflection effect and have orbital periods of $P<1\:\rm d$, making them vital tools for sdB studies due to the relative ease in identifying them.  The prototypical HW Vir is an sdB and M dwarf (dM) binary. A few systems containing a brown dwarf (BD) have also been discovered \citep[e.g.,][]{Schaffenroth2014}.  

The Eclipsing Reflection Effect Binaries from Optical Surveys (EREBOS) project \citep{Schaffenroth2019} is an effort to increase the sample of known HW Vir systems and to measure orbital, atmospheric, and fundamental parameters of those binaries. It is especially interested in finding the lower-mass limit of an object able to remove the envelope in a CE phase and survive this phase in order to investigate the effect that sub-stellar companions have on the late stages of stellar evolution. Moreover, this project aims at studying post-CE systems spanning the entire range of periods and companion masses for these systems.  For a better understanding of the poorly understood common envelope phase, see \cite{ivanova2013}. Until recently, the number of HW Vir binaries with known fundamental parameters was relatively small at 18 total systems. The EREBOS project dramatically increased this number by inspecting light curves from the Optical Gravitational Lensing Experiment \citep[OGLE;][]{OGLE2013,OGLE2015} and Asteroid Terrestrial-impact Last Alert System \citep[ATLAS;][]{ATLAS2018} surveys, finding over 150 new HW Vir candidates \citep{Schaffenroth2019}. With an extensive spectroscopic and photometric follow up campaign we will dramatically increase the number of systems with robust solutions.

Despite this unprecedented increase, HW Virs still represent only a small fraction of the sdB population.  Given the typical radii of both components, these systems have to be relatively edge--on to show any eclipse.  For example, the smallest, grazing eclipses occur in systems such ASAS 102322$-$3737 \citep{Schaffenroth2013}, a sdB+dM, at $i=65.9^{\circ}$; however, inclinations do range up to perfectly edge-on systems such as AA Dor \citep{Kilkenny1978}, a sdOB+dM/BD.  One 
HW Vir system, Konkoly J064029.1+385652.2 \citep{Derekas2015}, is an sdO+dM binary that even shows a total eclipse due to a relatively small ($R=0.096\,R_{\odot}$) sdO being in a nearly edge on ($i=87.11$) orbit with an inflated dM. Total eclipses are sometimes seen in WD+dM binaries such as NN Ser \citep{Parsons2010}, where a high inclination angle allows the dM to completely block the smaller WD along our line of sight. Due to the similarity in size between typical sdBs and dMs, even edge on systems struggle to achieve geometries sufficient to produce a total eclipse.

Here we present system parameters for the first deeply eclipsing sdB+dM system \textit{Gaia} DR2 6097540197980557440, which exhibits an
eclipse in excess of $\sim$5 magnitudes in the optical.  We discovered \target\ in the course of the EREBOS project while obtaining follow--up observations of known HW Virs using the Goodman spectrograph \citep{Goodman} on the 4.1--m Southern Astrophysical Research (SOAR). In \S \ref{sec:discovery} we describe the initial observations leading to its discovery. In \S \ref{sec:spectroscopy} we present time--series spectroscopic observations as well as the radial velocities and atmospheric parameters derived from them. In \S \ref{sec:photometry} we present multi--color, time--series photometric observations and the details of our light curve modeling solution. \S \ref{sec:parameters} presents system parameters derived from the best--fitting light curve modeling solution. In \S \ref{sec:discussion}, we discuss how the system compares to the EREBOS sample as well as potential follow-up studies.  Finally, we summarize our work in \S \ref{sec:summary}.

\section{\textbf{Discovery Run}} \label{sec:discovery}

\begin{figure*}[t]
    \centering
    \subfloat{
            \includegraphics[width=0.185\textwidth]{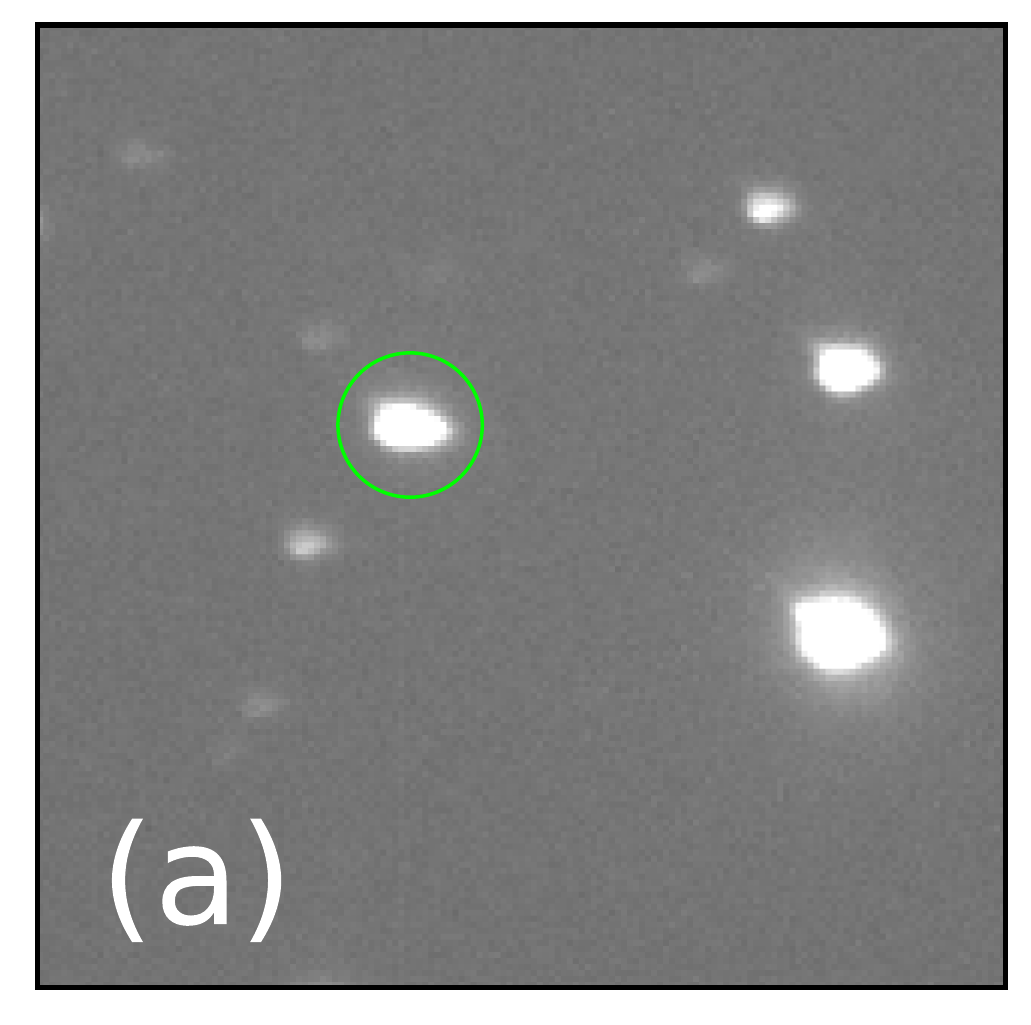}
            }
    \subfloat{
            \includegraphics[width=0.185\textwidth]{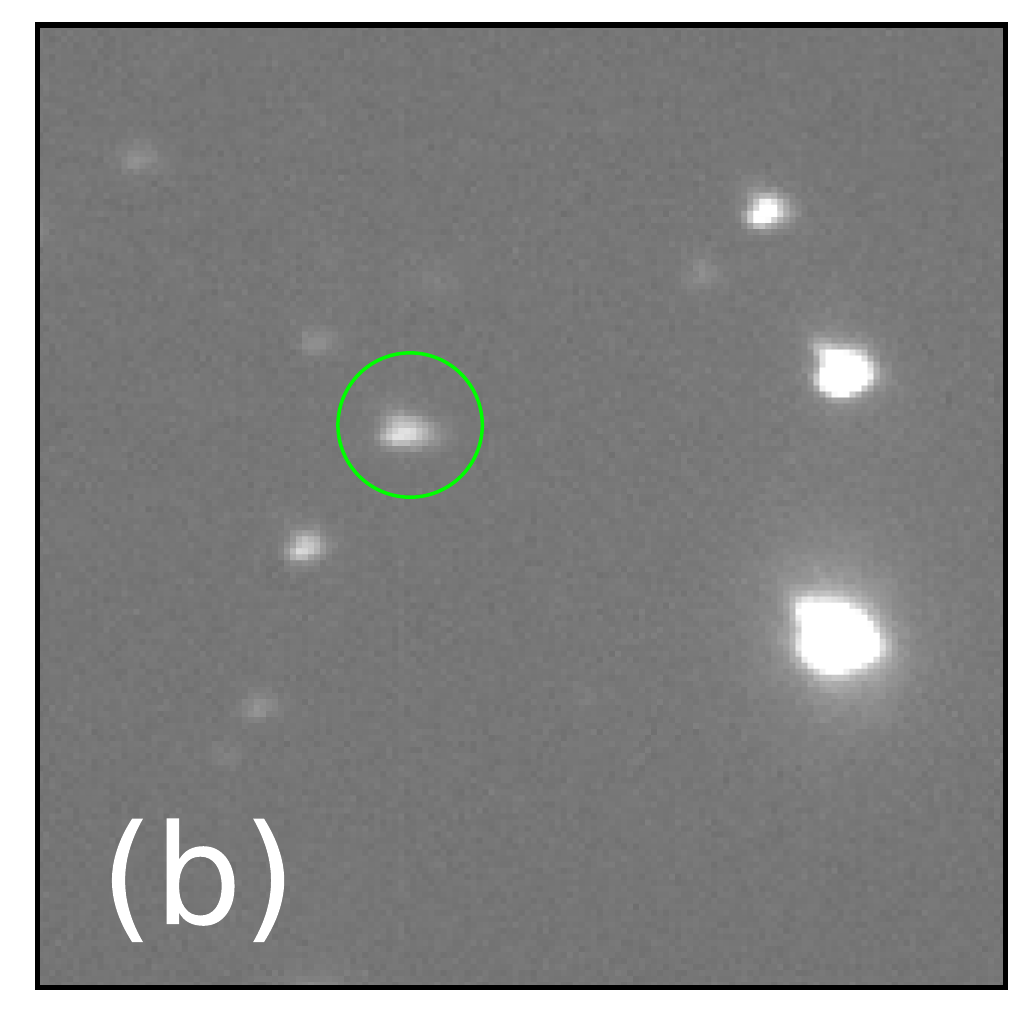}
            }
    \subfloat{
            \includegraphics[width=0.185\textwidth]{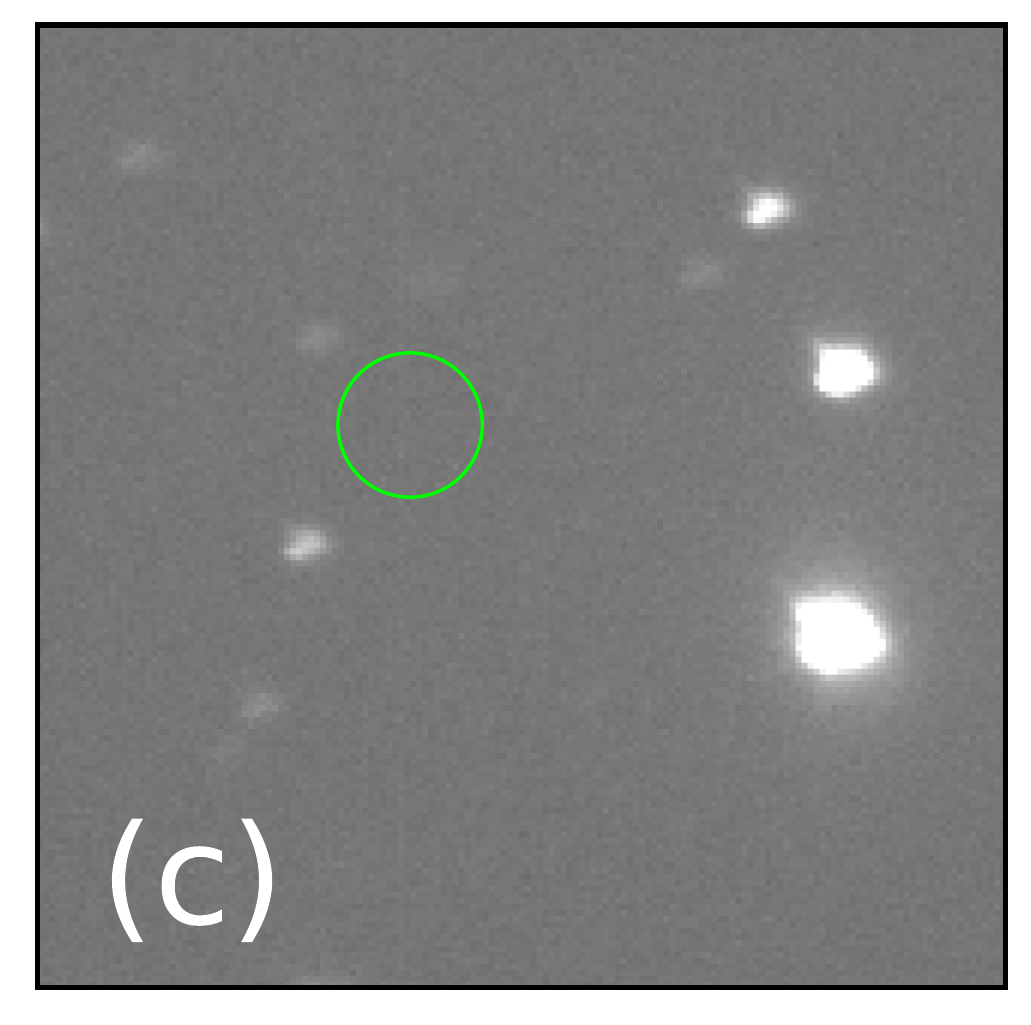}
            }
    \subfloat{
            \includegraphics[width=0.185\textwidth]{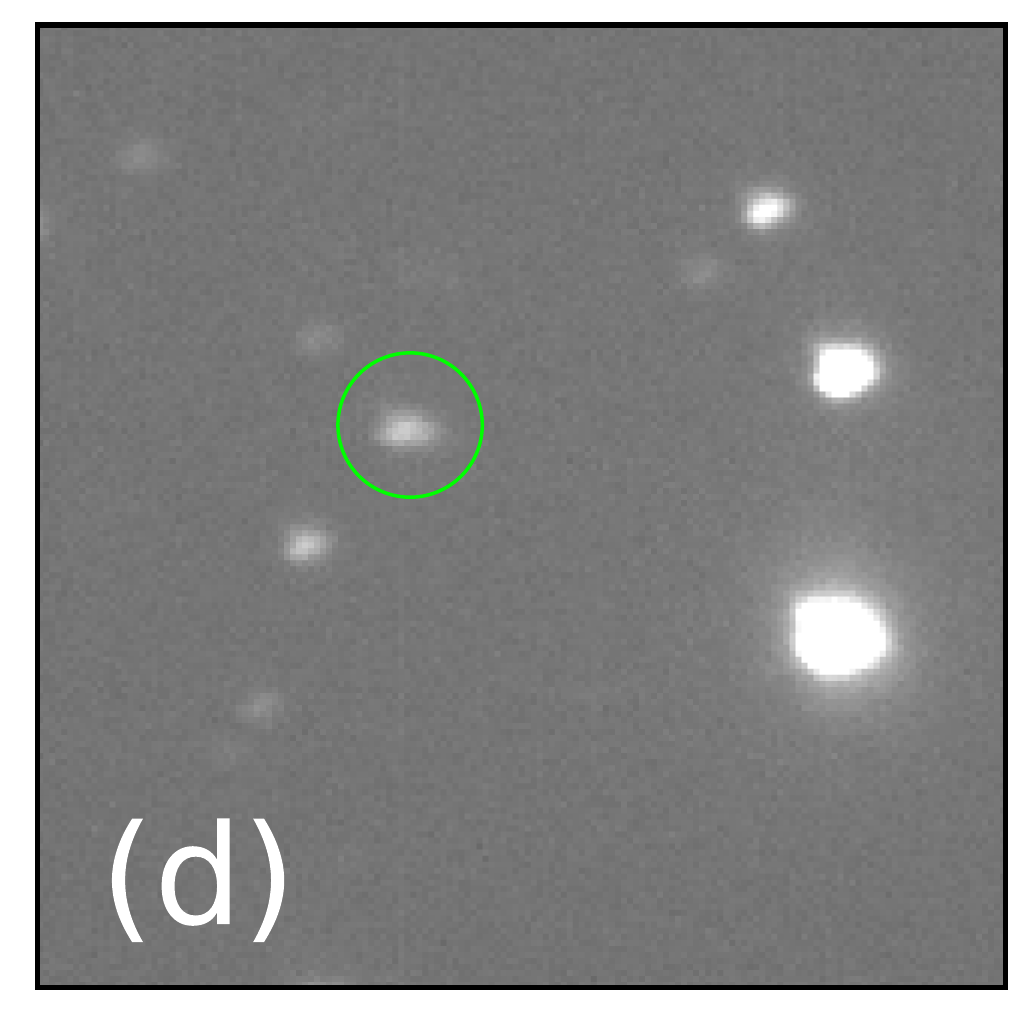}
            }
    \subfloat{
            \includegraphics[width=0.185\textwidth]{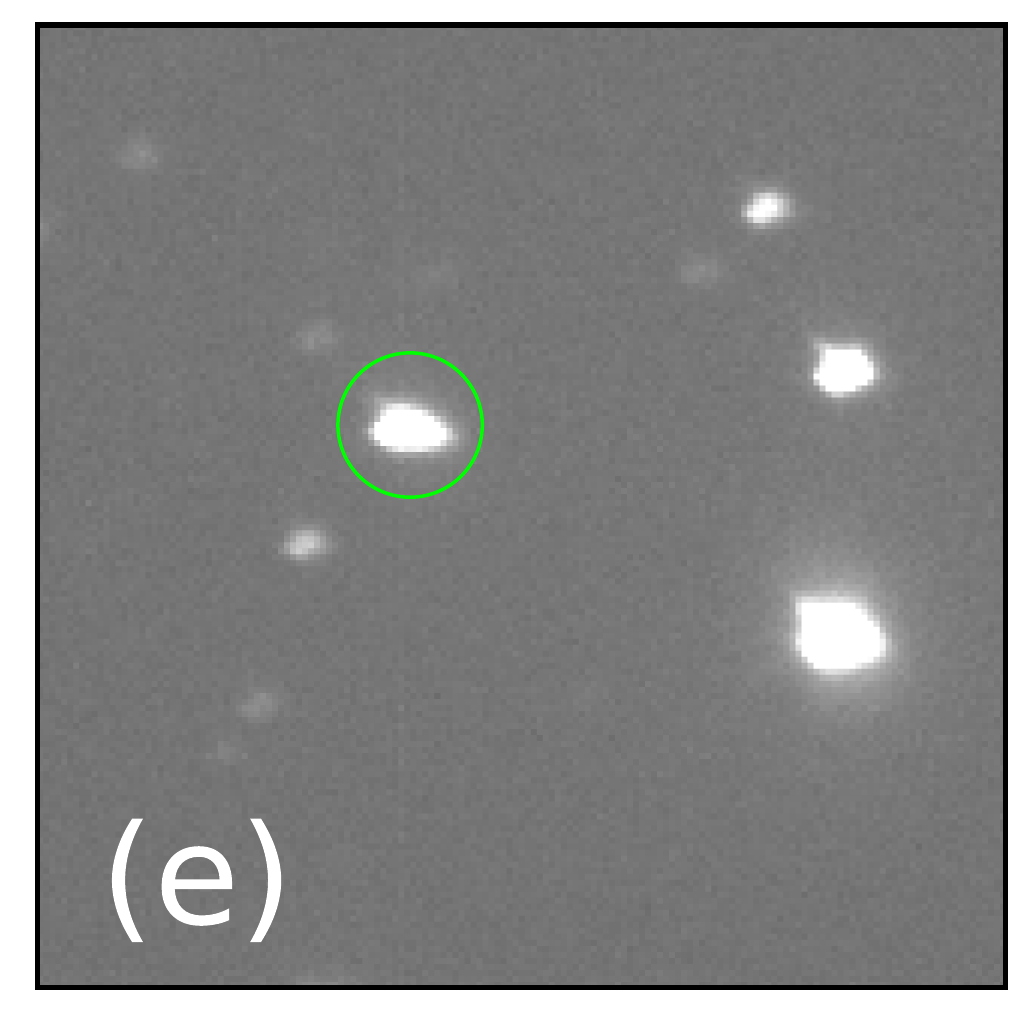}
            }\\
    \subfloat{\includegraphics[width=0.98\textwidth]{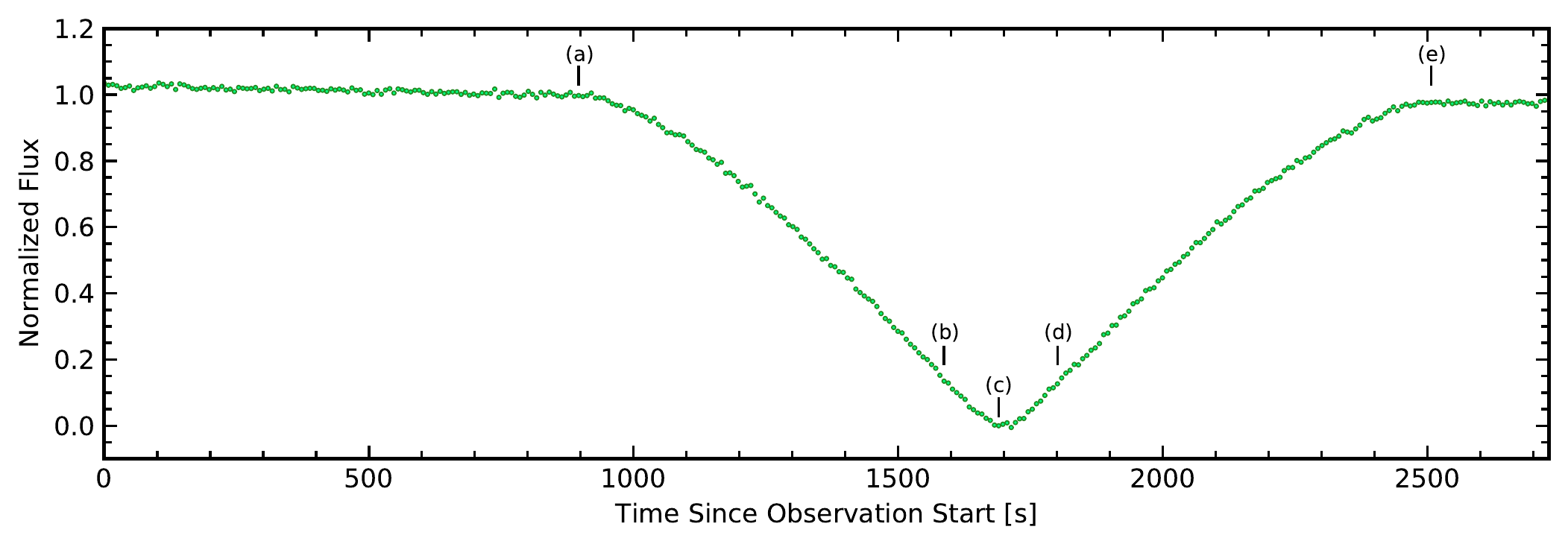}}
            
    \caption{Discovery data for \target\ from SOAR/Goodman. \textit{Top:} Raw Johnson $V$ filter frames from the discovery data set obtained on 2019 June 9.  We highlight five select frames corresponding to the marked locations on the light curve in the bottom panel.  These frames represent phases (a) just prior to ingress, (b) shortly before the systems drops below detection limits, (c) during primary eclipse totality, (d) shortly after the system returns above detection limits, and (e) just after egress. \textit{Bottom:} The corresponding light curve of \target\ in the Johnson $V$ filter.}
    \label{fig:frames}
\end{figure*}

During a small amount of down time between SOAR/Goodman observations of EREBOS targets on 2019 June 9, we discovered \target\ with approximately 45 minutes of time-series photometry using a Johnson $V$ filter.  We had previously identified \target\ as a strong candidate variable hot subdwarf from its anomalously high \textit{Gaia} DR2 photometric uncertainty, and its inclusion in the \citet{Geier2019} catalog of candidate hot subdwarf stars (see Barlow et al. in prep, \cite{Guidry2020} for details). We unwittingly began observing just before primary eclipse and, upon noticing the star disappear from the raw image frames\footnote{At this moment, Stephen Walser, who was monitoring the frames as they came in, apologetically informed us he had `lost our star.'} (shown in Figure \ref{fig:frames}), continued observing long enough to safely capture egress. Using the processes described in \textsection\ref{sec:LCobservations}, we constructed a light curve and determined that the primary eclipse lasted approximately 25 minutes.   The shape of the  eclipse stuck out to us immediately as being different than in other HW Vir binaries. Whereas the ingress and egress segments of most primary eclipses have {\em positive} second time derivatives (i.e. concave up), \target's second derivatives are {\em negative} during ingress and egress (i.e. concave down). This can only be explained by the geometry of a nearly perfectly edge--on eclipse, so we were eager to obtain photometry and spectroscopy over the full orbit to solve for all system parameters.

Unfortunately, we were unable to determine a precise orbital period for the system using our exploratory time--series photometry.  However, \target\ was also observed by TESS in Sector 11 through full-frame image (FFI), 30--min cadence observations.  The data were downloaded from the Mikulski Archive for Space
Telescopes (MAST) web portal, and the \texttt{lightkurve} \citep{lightkurve} Python package was used to extract time-series photometry from the FFIs. A Lomb-Scargle periodogram \citep{LombScargle1982} was computed and yielded an initial estimate of the system's orbital period of $P=3.0614\:\rm hr$.  This estimate helped guide subsequent observations.

\section{\textbf{Time--Series Spectroscopy}} \label{sec:spectroscopy}

\subsection{Observations \& Reductions} \label{sec:RVobservations}
We obtained 53 spectra using SOAR/Goodman on 2019 July 25 and 46 spectra on 2020 February 17, both in an uninterrupted series of back--to--back exposures.  Each of these data sets covered roughly 75\% of the $\sim$3--hr orbital period.  We used the $0.8''$ long slit, 2$\times$2 binning, and the 930 $\rm mm^{-1}$ VPH grating (0.84\,\AA\, per binned pixel dispersion), giving us average spectral resolutions of 2.38\,\AA\, and 2.04\,\AA\, over the wavelength range 3600-5300\,\AA\, for the 2019 and 2020 data, respectively. We note that the spectral resolutions are different despite using the same instrumental configuration due to the camera--collimator focus values not being set to their optimal values during the 2019 observations. On both observing nights, we aligned the slit axis to a position angle of $278.3^{\circ}$ E of N in order to place a bright star\footnote{Gaia DR2 6097528446950034944} $23.5''$ away on the slit and monitor any drifts in the wavelength solution due to instrumental flexure.  Individual spectra in each series were integrated for 120--s, yielding an average S/N of $\sim$30 per resolution element.  We also obtained spectra of FeAr lamps immediately following each series for wavelength calibration purposes.

Reduction of the frames was carried out using the \texttt{ccdproc} task in \texttt{IRAF} \citep{IRAF1986,IRAF1993}. After  bias-subtracting and flat-fielding all spectral images, we ran the \texttt{apall} task to extract a one-dimensional spectrum for each frame and remove a fit to the sky background.  For the 2020 data, a wavelength solution was generated from the FeAr lamp spectra and applied to all individual spectra. We note that slow drifts in the wavelength solution over the course of the series are expected due to instrumental flexure, and thus the FeAr wavelength solution does not provide an accurate zero--point --- only an accurate dispersion solution. For the 2019 data, an intermittent issue with the FeAr lamp prevented us from obtaining an accurate dispersion solution with it. Instead, we created a self template from the combined 2019 spectra and use the Balmer and He\,{\sc I} lines to determine the wavelength solution. Once again, this only provides a dispersion solution and not an absolute RV zero point. Consequently, we are unable to report on the binary's systemic velocity. The spectrum of 
\target , shown in Figure \ref{fig:atmospheric_fits}, is dominated by strong H Balmer absorption features and weaker He I lines (4026, 4471, 4921, 5015\,\AA). The absence of the He II 4686\,\AA\, line rules out an sdOB classification. 

\begin{figure*}
        \centering
        \subfloat{
            \includegraphics[width=0.49\textwidth]{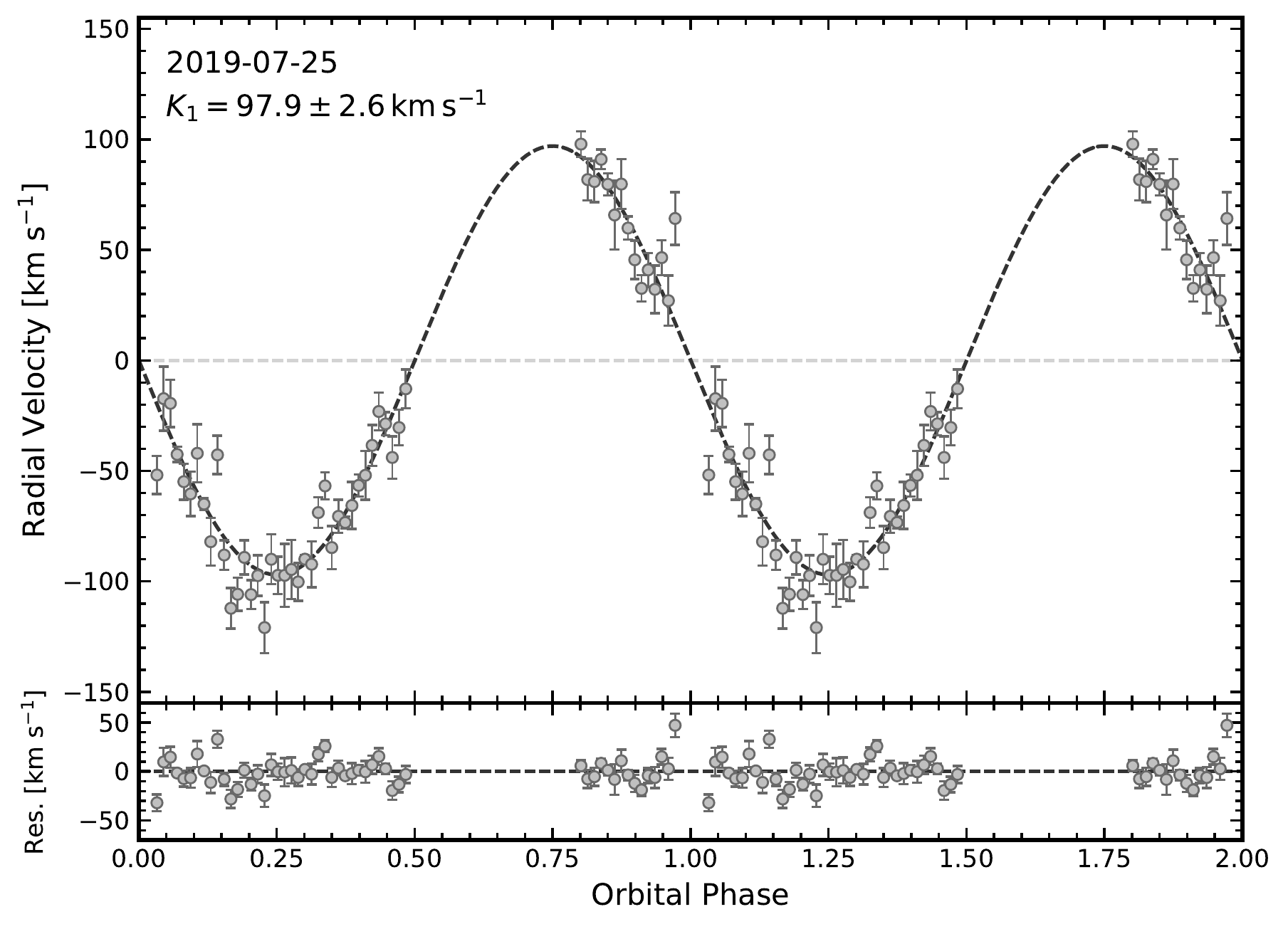}
            \label{fig:RV2019}
        }
        \subfloat{
            \includegraphics[width=0.49\textwidth]{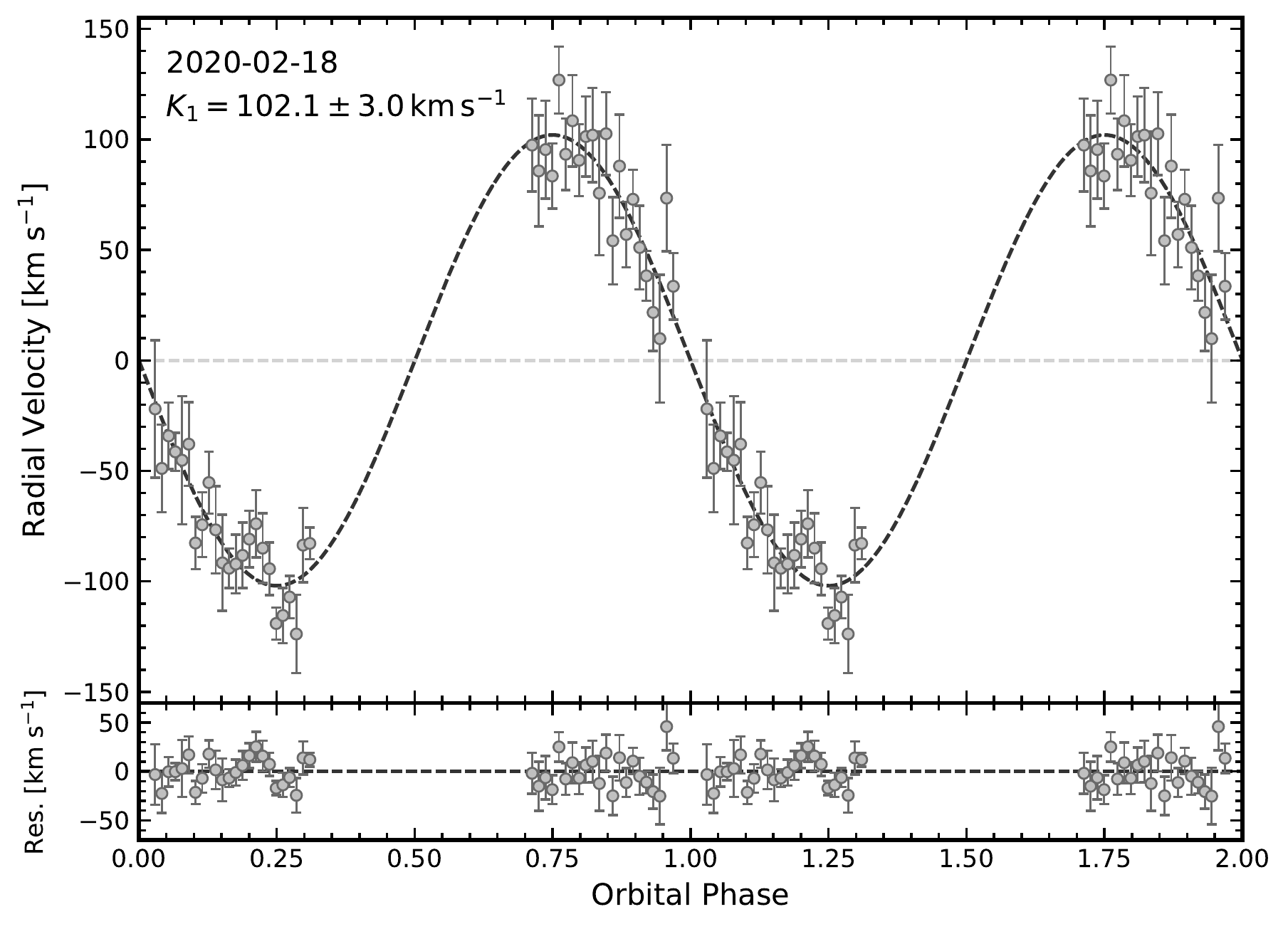}
            \label{fig:RV2020}
        }

        \caption{Radial velocity curves constructed from spectra obtained with SOAR/Goodman, plotted twice for better visualization.  \textit{Left:} Best fitting model for the data taken in 2019.  \textit{Right:} Best fitting model for the data taken in 2020.  Both of these solutions agree within the error bars with the weighted average $K_{\rm sdB}=100.0\pm2.0\rm\:km\,s^{-1}$.}
        \label{fig:RV_curves}
\end{figure*}

\subsection{Radial Velocity Curve} \label{sec:RVcurve}

Radial velocity (RV) shifts were determined from non--linear, least--squares fits of Gaussian profiles to the sdB H Balmer lines, which was carried out using the \texttt{curve\_fit} function in the Python package \texttt{scipy} \citep{2020SciPy-NMeth}.  The He I profiles were too noisy in individual spectra for this purpose. In order to correct for drifts in the wavelength solution (and thus drifts in the RVs) due to instrumental flexure during the observations, we also measured the relative velocity shifts of the absorption features of the second star on the slit. This object displayed spectral features consistent with a G/K--type star, and so we used the \texttt{crosscorrRV} function in the \texttt{PyAstronomy} library \citep{pya} to measure velocity shifts via cross correlation. The second star's RV curves
revealed gradual, nearly--linear shifts on the order of $\sim$75 km s$^{-1}$ over $\sim$2 hours, in both the 2019 and 2020 data sets. The magnitude and pattern of these shifts --- slightly different on the two nights --- were consistent with expectations given the target's RA, DEC, average hour angle during each run, and associated Nasmyth cage rotations. We are confident they are due to instrumental flexure and not intrinsic RV variations of the second star on the slit. To remove this flexure drift from the target RV curves, we fitted low--order polynomials to the comparison star's RV curves and subtracted this fit from the raw target star RV curves. The resulting RV curves are shown in Figure \ref{fig:RV_curves}.  

In order to determine the RV semi--amplitude of the sdB ($K_{\rm sdB}$), we fitted sine waves to each of the data sets separately, with the orbital period and phase fixed to the values described in \textsection\ref{sec:LC_modeling}. From the 2019 July 25 data, we find  $K_{\rm sdB}=97.9\pm2.6\rm\:km\,s^{-1}$, and from the 2020 February 17 data, we derive $K_{\rm sdB}=102.1\pm3.0\rm\:km\,s^{-1}$.  These results agree within their 1-$\sigma$ uncertainties, and we adopt as our final RV semi-amplitude their weighted average:  $K_{\rm sdB}=100.0\pm2.0\rm\:km\,s^{-1}$. The residuals in the bottom panels of Figure \ref{fig:RV_curves} are consistent with random noise and show the data are consistent with a circular orbit, as expected for post-common-envelope HW Vir binaries.

\subsection{Atmospheric Parameters} \label{sec:atomphericparams}

For use in the spectroscopic analysis, model spectra are computed following the so-called hybrid approach \citep{2006BaltA..15..107P,2006A&A...445.1099P,2008A&A...481..199N}. In this approach, deviations from local thermodynamic equilibrium (LTE) can be treated very efficiently using a combination of updated versions of the {\sc Atlas12} \citep{1996ASPC..108..160K}, {\sc Detail} \citep{1981PhDT.......113G,detailsurface2}, and {\sc Surface} \citep{1981PhDT.......113G,detailsurface2} codes. The {\sc Atlas12} code,  for which the mean metallicity for hot sdB according to \cite{2013MNRAS.434.1920N} is used here, is initially used to compute the temperature/density structure of a line-blanketed, plane-parallel, and chemically homogeneous atmosphere in hydrostatic and radiative equilibrium. This LTE atmosphere is then used as input for the {\sc Detail} code, which solves the coupled radiative transfer and statistical equilibrium equations to obtain occupation numbers in NLTE for hydrogen and helium. Finally, the {\sc Surface} code is used to compute the final synthetic spectrum using the atmosphere from {\sc Atlas12} and the occupation numbers from {\sc Detail} as well as more sophisticated line-broadening data. Also taken into consideration are the recent improvements to all three codes \citep{2018A&A...615L...5I} concerning NLTE effects on the atmospheric structure, the implementation of the occupation probability formalism \citep{1994A&A...282..151H} for hydrogen and neutral helium, and new Stark broadening tables for hydrogen \citep{2009ApJ...696.1755T} and neutral helium \citep{1997ApJS..108..559B}. The application of these models to sdBs is also shown in \citet{Schaffenroth2020}.

The observed spectra are matched to the model grid by $\chi^2$ minimization as 
described by \citet{1994ApJ...432..351S} as implemented by 
\citet{1999ApJ...517..399N}. We use six H Balmer lines and four He\,{\sc I} lines. H\,$\epsilon$ is excluded  because of contamination by interstellar Ca\,{\sc II}. 
Since the binary orbit is so tight, tidal forces probably have spun up the sdB star, which causes extra line broadening. However, the resolution of the spectra is insufficient to measure the projected rotational velocity v\,$\sin\,i$. We assume that the rotation of the sdB is tidally locked to the binary orbit and convolve the model spectrum with a rotational broadening profile with a corresponding v\,$\sin i$=87\,km\,s$^{-1}$ in the fitting procedure.  

Previous studies have shown that some sdBs with reflection effects have atmospheric parameters that can vary with phase when analyzing spectra of sufficiently high S/N taken at different phases of the orbit \citep[e.g.,][]{Schaffenroth2013,Schaffenroth2014b}.  These variations can be explained by the companion's phase-variable contributions to the spectrum from only the reflection effect, causing apparent variations of order 
$1000-1500$ K and 0.1 dex in the sdB temperature and surface gravity, respectively. 

\begin{figure*}
    \centering
    \subfloat{
            \includegraphics[width=0.50\textwidth]{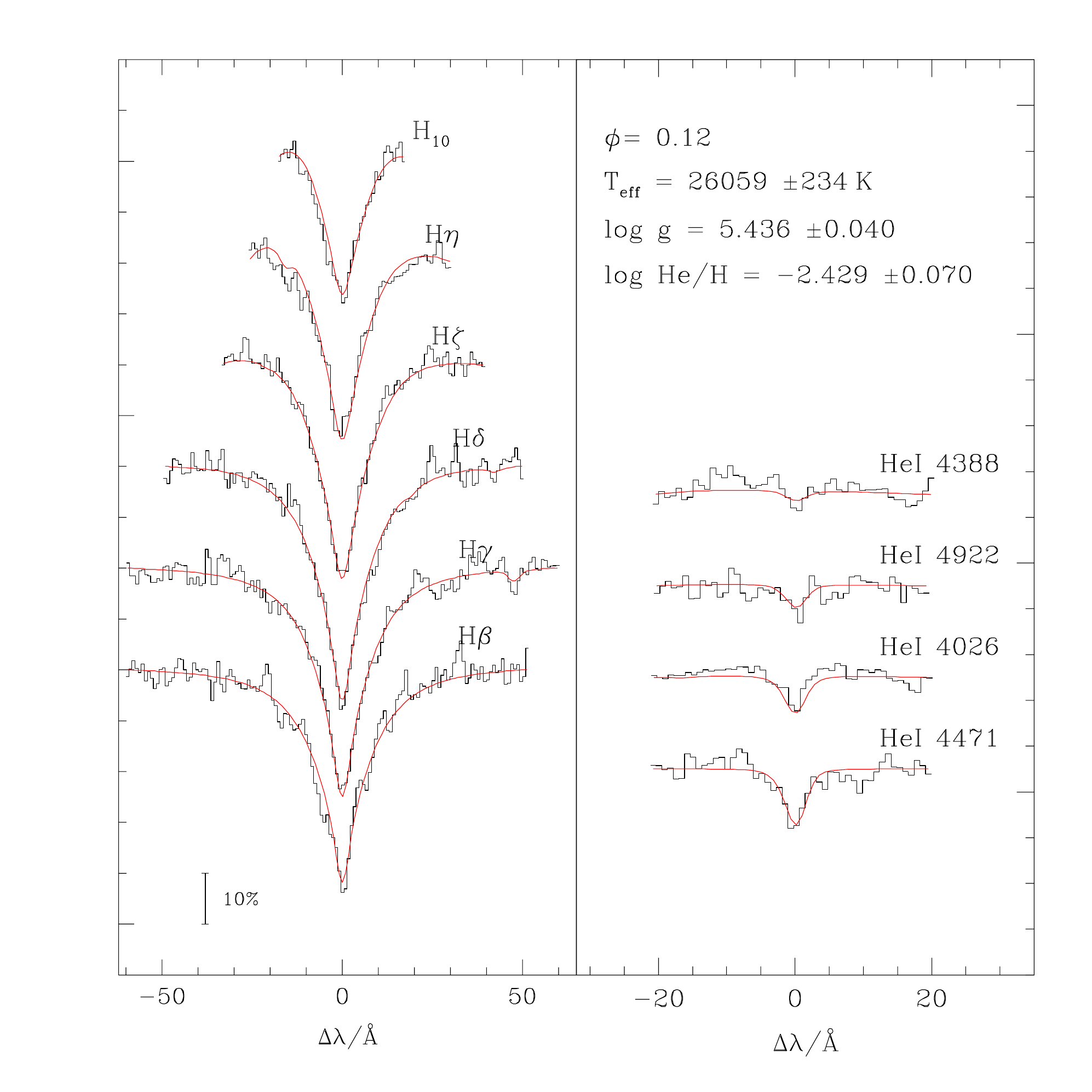}
            }
    \subfloat{
            \includegraphics[width=0.50\textwidth]{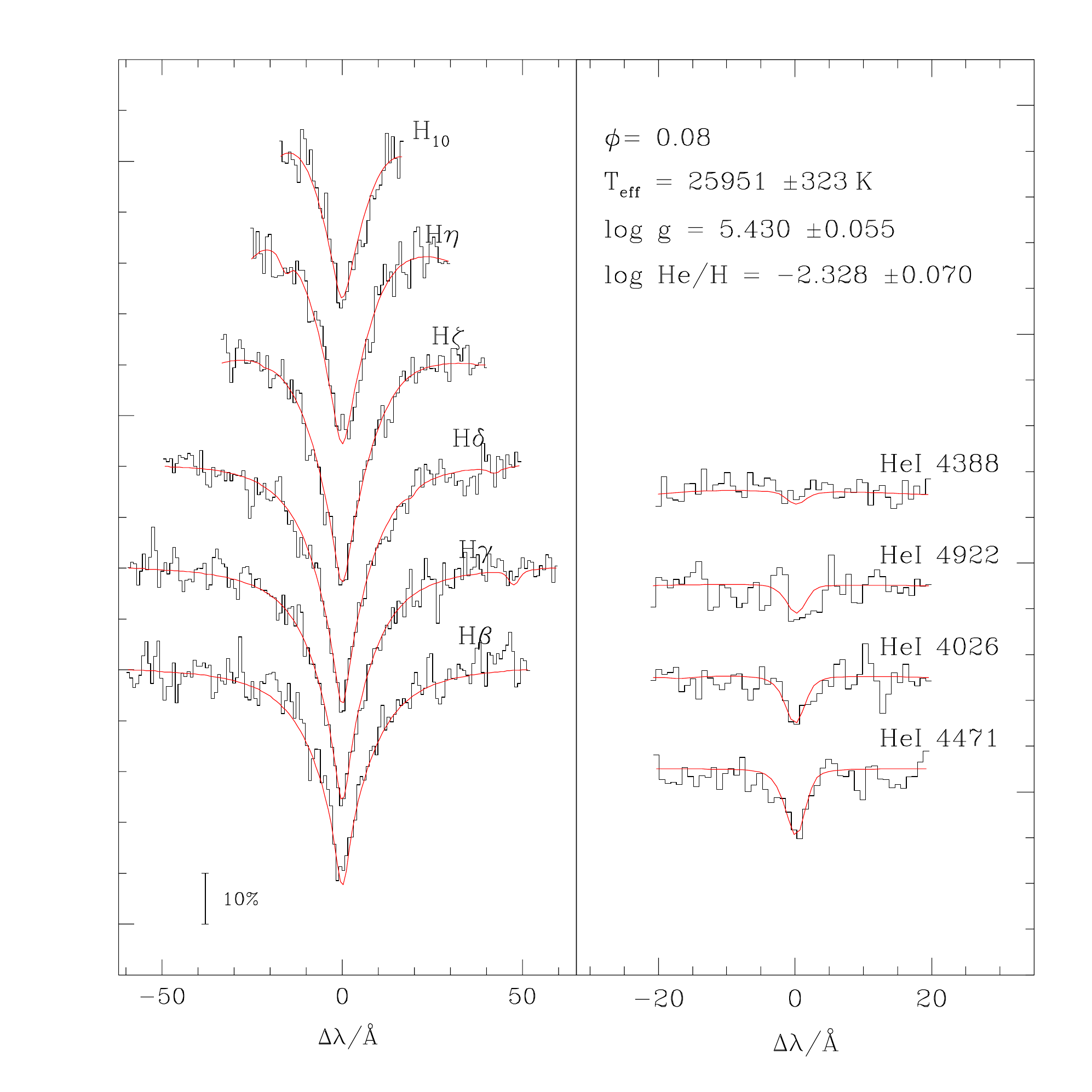}
            } \\
    \caption{Line fits to the hydrogen Balmer and neutral helium lines in individual SOAR/Goodman spectra from 2019 (left hand panel) and 2020 (right hand panel).  Listed in the upper right of each panel is the orbital phase and the resulting set of best fitting atmospheric parameters.}
    \label{fig:atmospheric_fits}
\end{figure*}

\begin{figure}
    \centering
    \includegraphics[width=\linewidth]{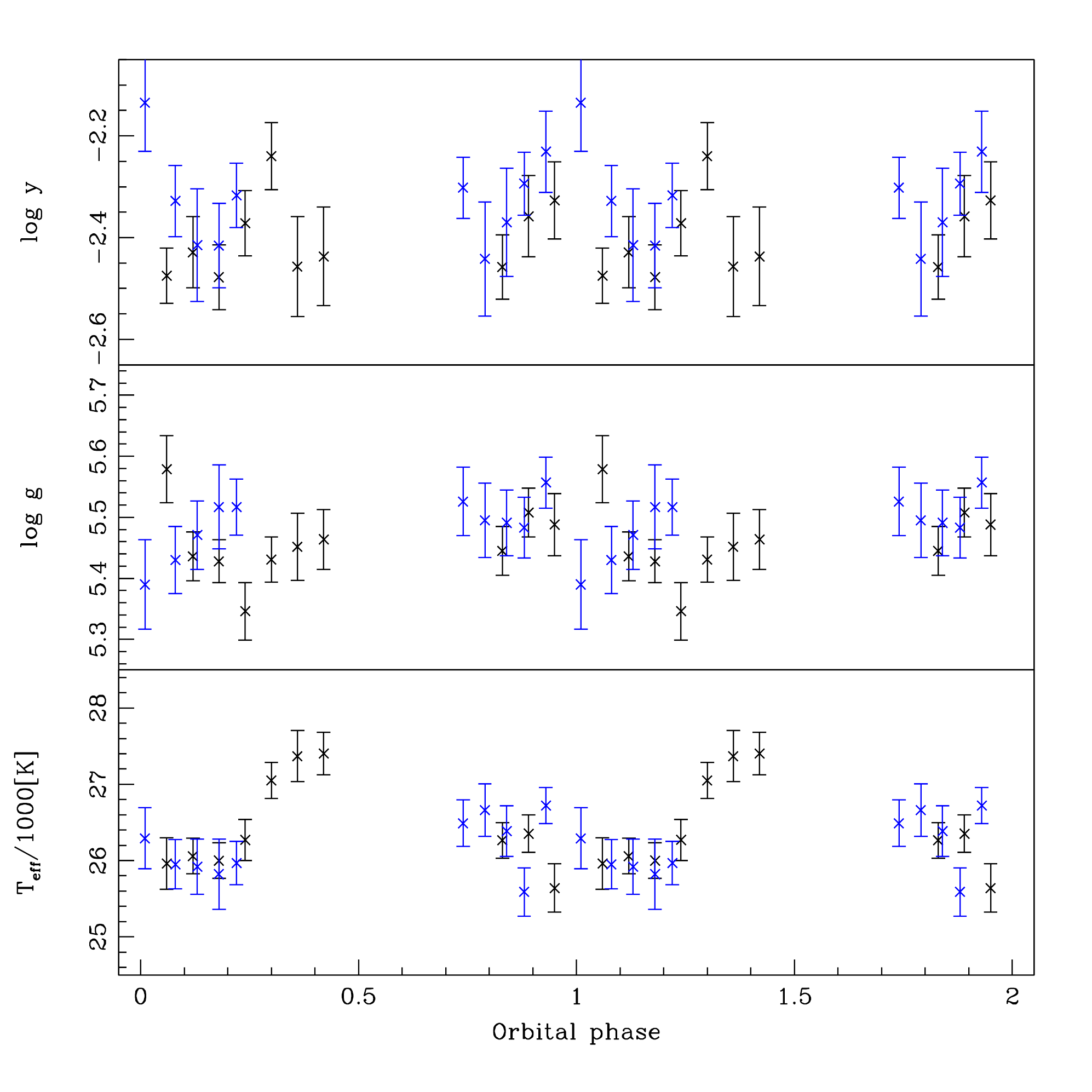}
    \caption{Apparent \teff\ (bottom), \logg\ (middle), \logy\ (top) variations with 1 $\sigma$ error bars as a function of the orbital phase. Results from spectra taken in 2019 are shown in black while those from 2020 in blue.
}    
\label{fig:ph_tgy}
\end{figure}

To account for any of these variations, we derived the atmospheric parameters from the single-radial velocity corrected spectra. Exemplary fits are shown for individual spectra from the 2019 and 2020 observing runs for similar orbital phases in Fig.~\ref{fig:atmospheric_fits}. Results from both observing runs are consistent.
The variations of the atmospheric parameters, which are consistent with previous determinations, can be seen in Fig. \ref{fig:ph_tgy}. The effective temperature appears to increase slightly near the secondary eclipse. Any variations in the surface gravity or helium abundance remain below detection limits. In order to determine the atmospheric parameters of the sdB we averaged the parameters near the primary eclipse, where only the dark side of the companion is visible: $T_{\rm eff}=26100\pm400$K, $\log\left(g\right)=5.50\pm0.07$, and $\log\left(y\right)=-2.32\pm0.10$.

\section{\textbf{Time--Series Photometry}} 
\label{sec:photometry}

\subsection{Observations \& Reductions} \label{sec:LCobservations}
Follow--up time--series photometry was obtained on 2020 February 18 using SOAR/Goodman in imaging mode.  In an effort to obtain multi-color photometry for more precise modeling, the filter wheel was manually switched between the Johnson $B$ and $R$ filters every few minutes when not in primary or secondary eclipse, and every 30 seconds during eclipses. The integration time was fixed to 5 seconds for both filters in order to minimize dead time and errors associated with changing this value back and forth every few minutes. We used 2 $\times$ 2 binning and read out only a 350 $\times$ 175 binned pixel subset of the image to minimize the readout time between exposures. This relatively small field still provided several nearby comparison stars through which to track sky transparency variations. We achieved a duty cycle of roughly 54\% over the course of our observations, which covered a little more than one full orbital period. A more efficient duty cycle would have required either decreasing the subframe region further and sacrificing comparison stars, or increasing the exposure time and risk saturating the target and comparison stars.

\begin{figure*}
    \centering
    \includegraphics[width=\textwidth]{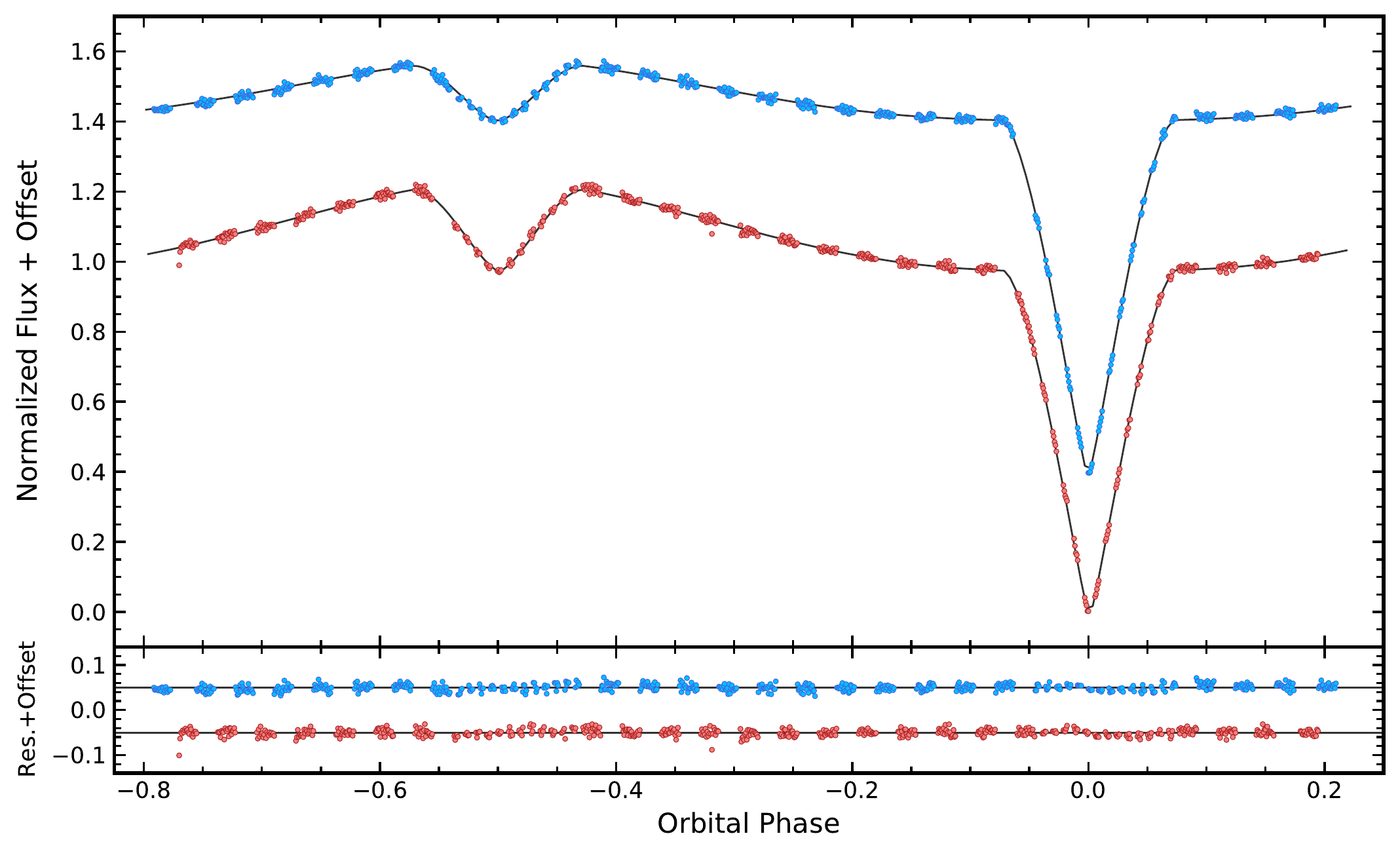}
    \caption{SOAR/Goodman light curves of \target\ in both the Johnson $B$ (blue points) and $R$ (red points) filters, along with their respective best-fitting models from \S\ref{sec:LC_modeling}. The $B$--filter light curve is offset by 0.4 for better visualization. Residuals are shown in the bottom panel with offsets of 0.05 and $-$0.05 for the $B$ and $R$ curves, respectively.}
    \label{fig:LC_models}
\end{figure*}

Reduction of the SOAR frames was once again carried out using the \texttt{ccdproc} procedure in \texttt{IRAF}.  Each raw image frame was first bias-subtracted and flat-fielded. We then extracted aperture photometry using a range of aperture sizes with a custom code utilizing the \texttt{photutils} \citep{Photutils} Python package. Sky counts were removed using sky annuli drawn around the apertures. Apertures were chosen to maximize the signal-to-noise ratio (S/N) in each light curve. This process was repeated on multiple nearby, bright comparison stars to remove sky transparency variations and flux-normalize the light curves.  Multiple cycles of observing are typically needed to remove airmass--related changes in the flux, therefore, any of these slight flux variations were not removed during the reduction process.  The resulting differential light curves are shown in Figure \ref{fig:LC_models} and used for modeling the binary. 

\subsection{Binary Light Curve Modeling}\label{sec:LC_modeling}
The \target\ light curve exhibits all the typical HW Vir features.  The amplitude of the reflection effect is noticeably stronger in the $R$ filter ($\sim30\%$) than in the $B$ filter ($\sim20\%$), and it is quite strong in general compared to other reflection effect systems.  Initially, this led us to believe that either the sdB was slightly hotter than in typical HW Virs, or the companion was slightly larger than usual. The deep primary eclipse, implying a nearly edge--on inclination, lent credence to the latter explanation. The shape of the eclipse itself sticks out amongst other HW Vir binaries. As mentioned in Section \ref{sec:discovery}, the ingress and egress segments of the primary eclipse have negative second derivatives (more V--shaped) instead of the more frequently observed positive second derivatives (more U--shaped). This implies the eclipse geometry is nearly perfectly edge--on and that the companion might be slightly larger than the primary. Secondary eclipses are also present in the light curve, during which the sdB is blocking irradiated light from the cool companion. Notably, the flux at the center of the secondary eclipse returns to its exact value immediately preceding and following ingress and egress, respectively --- further implying that the inclination must be nearly edge-on. 

To model the light curves, we use the code \texttt{LCURVE} \citep[for details, see Appendix A in][]{Copperwheat2010}.  In addition to recreating deep eclipses, \texttt{LCURVE} was designed for binaries with WDs and has been used to fit WD+dM systems exhibiting the reflection effect \citep[e.g.,][]{Parsons2010}; therefore, HW Vir binaries are naturally suited to be modeled in a similar fashion 
\citep[see][for an example and further details]{Schaffenroth2020}.
To form full solutions for these systems, there are many parameters that are not all independent, so we can greatly improve our ability to constrain each solution by fixing as many parameters as possible.  We fixed the sdB temperature to the value determined in our spectroscopic fit (described in \S\ref{sec:atomphericparams}).  We also fix the gravitational limb darkening coefficients to values expected of a primary with a radiative atmosphere \citep{vonZeipel1924} and a companion with a convective atmospheres \citep{Lucy1967} by calculating the resulting intensities using a blackbody approximation.  Then we adopted a quadratic limb darkening law for the primary using the values in the \citet{Claret2011} closest to the parameters derived in our spectroscopic fits. 

It is important to note that there is a large degeneracy in the light curve solutions of HW Vir binaries, even when fixing all of the above parameters.  The orbit is certainly almost circular, so each model is not sensitive to the mass ratio ($q$) of the system.  For this reason we calculated different solutions over a range of various, fixed mass ratios.  We then use a SIMPLEX algorithm \citep{press92} to vary parameters such as the inclination, both radii, the companion's temperature, albedo, and limb darkening, and even the period and primary eclipse time to help localize the solutions.  Additionally, we allow for linear trends due to airmass--related changes in flux over the course of the observations.

Next we tested the degeneracy of each light curve solution and determined the parameter errors by performing Markov-Chain Monte-Carlo (MCMC) computations using \texttt{emcee} \citep{Foreman-Mackey2013}.  We used the best-fit solution
from our SIMPLEX algorithm for initial values, and then we again varied the inclination angle, both radii, the limb darkening coefficient assuming a linear limb darkening law for the companion, and the companion's temperature and albedo for the mass ratio of our most probable solution (see \textsection\ref{sec:parameters}).  In all cases, the temperature of the companion is not well constrained as its fractional luminosity contribution to the system --- outside of the reflection effect --- is negligible. We therefore constrained the companion's temperature to the range $2500-3500\,$K (the expected range for the low-mass companion). The results and errors from our \texttt{emcee} run (shown in Table \ref{tab:lc_model_params}) then form the basis for our most probable solution.

\begin{figure}
    \centering
    \includegraphics[width=\linewidth]{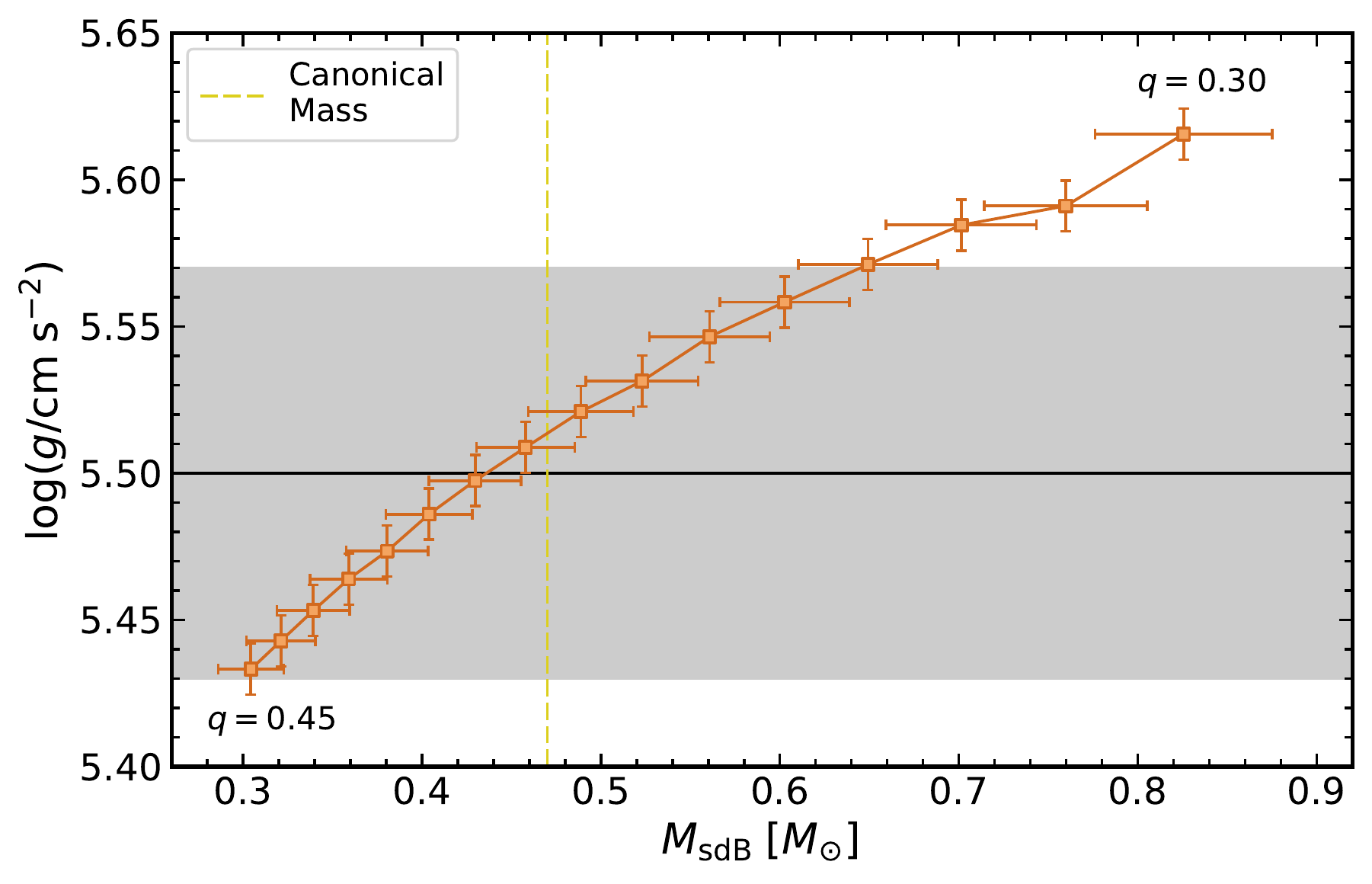}
    \caption{Photometric surface gravities plotted against their respective sdB masses for different mass ratio solutions ranging from q = 0.3 -- 0.45, in 0.01 increments.  The solid horizontal line (black) and shaded area represent the spectroscopically--derived $\log(g)$ and associated 1-$\sigma$ error, respectively.  The vertical dashed line (yellow) represents the canonical sdB mass of $0.47\,M_{\odot}$. The intersection of these two lines shows that our most probable solution is the one with a sdB mass just below the canonical mass.}
    \label{fig:logg-M}
\end{figure}

\begin{table*}
    \centering
    \small
    \caption{Parameters used to model the light curve for both the SOAR/Goodman $B$ and $R$ data.}
    
    \begin{tabular}{@{} lrrcl @{}}
    \toprule
    
    \multicolumn{1}{l}{\small{Parameter}} & 
    \multicolumn{1}{r}{\small{SOAR/Goodman -- $B$}} &
    \multicolumn{1}{r}{\small{SOAR/Goodman -- $R$}} & 
    \multicolumn{1}{c}{\small{Unit}} &
    \multicolumn{1}{l}{\small{Description}} \\
    
    \midrule
    \multicolumn{5}{c}{Fixed Parameters} \\
    \midrule
    
    $q\;(M_{\rm dM}/M_{\rm sdB})$ & 0.375 & 0.375 &  & mass ratio\\
    $P$ & $0.127037$ & $0.127037$ & d & orbital period \\ 
    $T_{\rm sdB}$ & 26100 & 26100 & K & primary temperature from spectroscopy \\
    $\textsl{g}_{1}$ & $0.25$ & $0.25$ &  & gravitational darkening exponent \\
    $\textsl{g}_{2}$ & $0.08$ & $0.08$ &  & gravitational darkening exponent \\
    $x_{1,a}$ & $0.097$ & $0.070$ & & primary linear limb darkening coefficient\\
    $x_{1,b}$ & $0.285$ & $0.222$ & & primary quadratic limb darkening coefficient\\

    \midrule
    \multicolumn{5}{c}{Adjusted Parameters} \\
    \midrule
    
    $i$ & $90^{+0.0}_{-0.3}$ & $90^{+0.0}_{-0.4}$ & $^{\circ}$ & inclination angle\\
    $x_{2, a}$ & $0.2992$ & $0.2734$ & & companion linear limb darkening coefficient\\ 
    $R_{\rm sdB}/a$ & $0.2180\pm0.0007$ & $0.2174\pm0.0007$ &  & primary radius \\
    $R_{\rm dM}/a$ & $0.2402\pm0.0006$ & $0.2407\pm0.0006$ &  & companion radius \\
    $T_{\rm dM}$ & $2800\pm500$ & $3100\pm500$ & K & companion temperature \\
    $A_{2}$ & $1.14\pm0.01$ & $1.4\pm0.01$ &  & companion albedo (absorb) \\
    $m$ & $0.00026\pm0.00001$ & $0.00026\pm0.00001$ &  & slope \\

    \bottomrule
    
    
    \end{tabular}
    
    \label{tab:lc_model_params}
\end{table*}

\subsection{Orbital Ephemeris}\label{sec:orbitephemeris}
To aid in future observations of \target, we have listed its orbital ephemeris ($T_0$, $P$) in Table \ref{tab:derived_params}.   We adopt the  orbital period from our best fit \texttt{emcee} solution in \textsection\ref{sec:LC_modeling}.  To construct an initial eclipse time ($T_0$), we fit inverted Gaussian profiles to both the $B$ and $R$ time-series data using \texttt{curve\_fit}.  We then adopt the weighted average of the central times from both filter series as our $T_0$ value.

\section{\textbf{System Parameters}} \label{sec:parameters}
In Figure \ref{fig:logg-M}, we plot the surface gravity and sdB mass for each of the potential solutions, and we compare the photometric surface gravities to our spectroscopically derived surface gravity.  

Based on the spectroscopic surface gravity, we get a consistent solution for an sdB mass of $0.3-0.64\,\rm M_\odot$. All possible solutions fit the light curve data nearly equally well; thus, we cannot claim a unique solution without additional data (e.g., velocity measurements from the dM). The most probable solution is the one with an sdB mass consistent with the canonical mass of $0.47\,M_{\odot}$.  The adopted best--fitting light curve solution and all relevant parameters are given in Table \ref{tab:lc_model_params}, and both of these best-fit models are shown together with their respective observations and residuals in Figure \ref{fig:LC_models}.  All possible solutions are given in Table \ref{tab:q_solutions} of the Appendix.

We compute the binary mass function for \target\ using the expression
\begin{equation}
    f = \frac{K_{\rm sdB}^{3}P}{2\pi G} = \frac{M_{\rm sdB}q^{3}\sin^{3}i}{\left(1+q\right)^{2}},
\end{equation}
finding $f=0.0132\pm0.0008 M_{\odot}$ using the period and sdB velocity semi-amplitude.  Combining this with the adopted mass ratio derived before, we find the sdB and dM masses to be $M_{\rm sdB}=0.47\pm 0.03\, M_{\odot}$ and $M_{\rm dM}=0.18\pm0.01\, M_{\odot}$, respectively.  Using Kepler's third law, we then find the orbital separation to be $a=0.921 \pm 0.018\, R_{\odot}$.  We also find $R_{\rm sdB}=0.199 \pm 0.004\, R_{\odot}$ and $R_{\rm dM}=0.222\pm 0.004\, R_{\odot}$.  Table \ref{tab:derived_params} gives an overview of the adopted parameters for \target.

\begin{figure}[h]
    \centering
    \includegraphics[width=\linewidth]{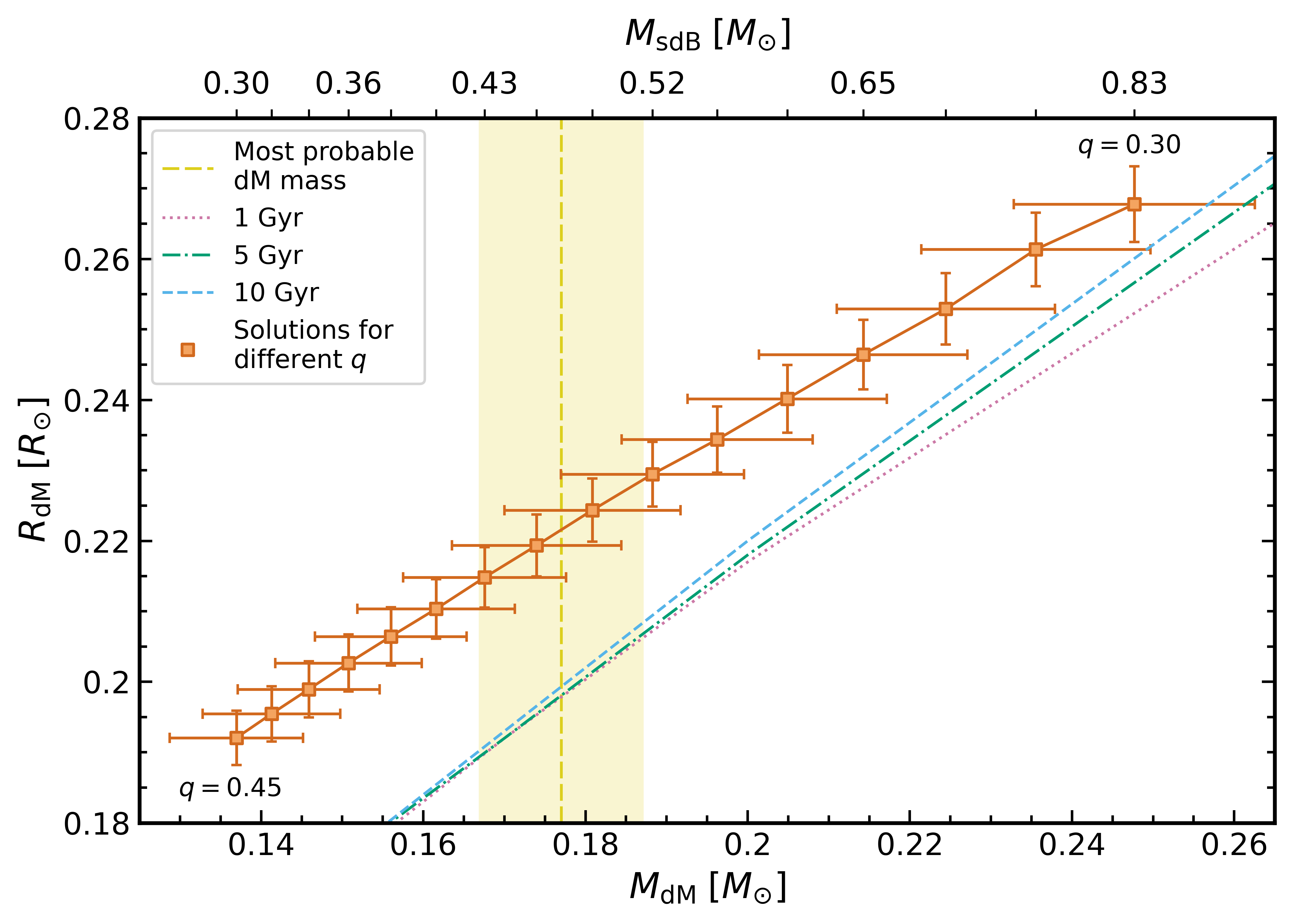}
    \caption{Mass--radius diagram for the dM companion illustrating the degeneracy in model solutions (orange squares). Theoretical mass-radius relations of low-mass stars \citep{Baraffe2015} for a 1 Gyr (dashed blue line), 5 Gyr (dash-dotted green line), and 10 Gyr (dotted pink line) are also included.  The vertical yellow line and shaded region represent the most probable dM mass and 1-$\sigma$ error, respectively, associated with the adopted $0.47\,M_{\odot}$ sdB solution.}
    \label{fig:M-R_relations}
\end{figure}

In Figure \ref{fig:M-R_relations}, we show each set of parameters for the companion and the theoretical mass-radius relations for low-mass main sequence stars from \citet{Baraffe2015} as an additional check. It is clear that each solution yields a companion radius that is inflated relative to what is predicted by theory, which is a trend commonly seen in close binaries with M dwarf components \citep{parsons2018}. For our most probable solution we get a companion inflation of about $\sim$13\%.

\begin{table}
    \centering
    \caption{Overview of derived parameters for \target\ that represent the most probable solution from the set of potential solutions.}
    
    \begin{tabular}{@{} lrr @{}}
    \toprule
    
    \multicolumn{1}{l}{\footnotesize{Parameter}} & \multicolumn{1}{c}{\footnotesize{Value}} & \multicolumn{1}{r}{\footnotesize{Unit}} \\
    
    \midrule
    \multicolumn{3}{c}{Basic Information} \\
    \midrule
    
    $\alpha^{a,b}$ & 213.577775581303 & deg \\
    $\delta^{a,b}$ & $-43.552249057309$ & deg \\
    $G^{a}$ & 16.358994 & mag \\
    $G_{\rm bp}-G_{\rm rp}\!^{a}$ & $-0.27529526$ & mag \\
    
    \midrule
    \multicolumn{3}{c}{System Properties} \\
    \midrule
    
    $T_0$ & $2458898.85724\pm0.00003$ & BJD \\
    $P$ & $0.127037\pm0.000001$ & d \\
    $i$ & $90^{+0.0}_{-0.3}$ & $^{\circ}$ \\
    $q$ & $0.375\pm0.003$ & \\  
    $a$ & $0.921\pm 0.018$ & $R_{\odot}$ \\
    
    \midrule
    \multicolumn{3}{c}{sdB Properties} \\
    \midrule
    
    $M_{\rm sdB}$ & $0.47\pm 0.03$ & $M_{\odot}$ \\
    $R_{\rm sdB}$ & $0.199\pm 0.004$ & $R_{\odot}$ \\
    $T_{\mathrm{eff}}$ & $26100\pm400$ & K \\
    $\log\left(g\right)$ & $5.50\pm0.07$ &  \\
    $\log\left(y\right)$ & $-2.32\pm0.10$ &  \\
    $K_{\rm sdB}$ & $100.0\pm2.0$ & $\mathrm{km\,s^{-1}}$ \\

    \midrule
    \multicolumn{3}{c}{dM Properties} \\
    \midrule
    
    $M_{\rm dM}$ & $0.177\pm0.010$ & $M_{\odot}$ \\
    $R_{\rm dM}$ & $0.222\pm 0.004$ & $R_{\odot}$ \\
    $T_{\mathrm{eff}}$ & $3000\pm500$ & K \\

    \bottomrule
    
    \multicolumn{3}{@{}l}{\textbf{\small{Notes}}}\\
    \multicolumn{3}{@{}l}{\footnotesize{$^{a}$ From \textit{Gaia} DR2 \citep{Gaia2018}}} \\
    \multicolumn{3}{@{}l}{\footnotesize{$^{b}$ Epoch J2015.5.}}
    
    \end{tabular}
    
    \label{tab:derived_params}
\end{table}

\section{\textbf{Discussion}} \label{sec:discussion}

Our analysis of \target\ represents the first EREBOS case study following the \citet{Schaffenroth2019} report of newly discovered sdB+dM systems.  With each additional system that is solved, EREBOS comes one step closer to achieving one of its goals to make statistical statements about a homogeneously selected population of close sdB systems.  While one new system by itself might not push the boundaries of key parameters in these studies, each system provides self-consistent feedback about the methodology used to study the overall population.  It is only through these self-consistent measures that EREBOS can eventually make statements regarding the effects stellar and sub-stellar companions have on the late stages of stellar evolution.

Our light curve and atmospheric modeling solutions imply \target\ is a fairly typical sdB+dM system, aside from the chance alignment of its orbital plane nearly perfectly along our line--of--sight.  The peak of the EREBOS orbital period distribution for both new and previously--published systems from \citet{Schaffenroth2019} is at $P=0.1\,$d, meaning that \target\  falls at the typical period for HW Virs.  
The most probable solution is a sdB with a mass of the canonical mass $M_{\rm sdB}=0.47\pm 0.03\,M_{\odot}$.
Additionally, our derived $\log\left(g\right)$ and $\log\left(y\right)$ values are also fairly typical of sdBs in HW Virs, but it is worth noting that our $T_{\rm eff}$ value is slightly lower than is typically found \citep[for comparison, see Fig. 6 in][]{Schaffenroth2019}.

There are also noteworthy aspects of the system that are somewhat atypical among HW Virs, namely the derived companion mass and sdB velocity semi-amplitude.  
The companion mass is tied for the most massive in an HW Vir binary, along with that of Konkoly J064029.1+385652.2 -- also a deeply--eclipsing HW Vir--type (sdO+dM) binary. \target\ has an orbital period that is $\sim$1.5 hr shorter than Konkoly J064029.1+385652.2 and will one day evolve into a more rapid analog of Konkoly J064029.1+385652.2 when the sdB evolves into an sdO after the He in the core is exhausted and then, inevitably, into a WD.
The sdB semi-amplitude we derive from the two sets of RV data make \target\ the fastest line--of--sight sdB velocity semi-amplitude reported to date for an HW Vir binary.

The most striking aspect of \target\ is the total eclipse of the sdB by its companion.  Due to this system being relatively bright ($G\sim16.4$ mag), a large eclipse depth means future eclipse timing (O$-$C) analyses to search for changes in the orbital period ($\dot{P}$) and even R{\o}mer delay studies should be possible using telescopes with a variety of aperture sizes \citep[e.g.,][]{Barlow2012}.  Additionally, \target\ will be observed at 2-min cadence in Sector 38 of TESS Cycle 3 through the Guest Investigator program (proposal \#G03221), providing space-quality data spanning 27 d of observations.  This is a unique opportunity to explore a relatively novel parameter space with one of the most accurate astrophysical clocks known \citep[e.g.,][]{Kilkenny2014}.

\section{\textbf{Summary}} \label{sec:summary}
We have presented photometric and spectroscopic observations of \textit{Gaia} DR2 6097540197980557440, the first deeply eclipsing sdB+dM binary.  Other than the remarkably striking nature of the eclipse, the system is a rather typical sdB+dM system.  We find an orbital period of $P=0.127037$ d and an sdB velocity semi-amplitude of $K_{\rm sdB}=100.0\rm\:km\,s^{-1}$ which combined with the most probable light curve solution yields masses of $M_{\rm sdB}=0.47\,M_{\odot}$ and $M_{\rm dM}=0.18\,M_{\odot}$, respectively.  This gives a radius of $R_{\rm dM}=0.222\, R_{\odot}$ for the companion, which is slightly inflated relative to theoretical mass--radius relationships of low-mass main sequence stars.  
\target\ represents the first HW Vir solved as part of the EREBOS project. Eventual solutions for the more than 100 new HW Vir binaries uncovered by EREBOS will help improve our understanding of the common envelope channel leading to sdBs, and help determine the effects nearby low-mass stellar and substellar objects can have on stars climbing the giant branch. 

\section*{\textbf{Acknowledgements}}
K.A.C., B.N.B., and S.W. acknowledge the support of the National Science Foundation under grant AST-1812874. BB also acknowledges partial support of this work by NASA through the TESS Guest Investigator program, under grant 80NSSC19K1720.
V.S. is supported by the Deutsche Forschungsgemeinschaft (DFG) through grant GE 2506/9-1.  A.I. acknowledges funding by the DFG through grant HE1356/70-1.

This research draws upon data provided by NOIRLab 2019A-0255 and 2020A-0268; PI B. Barlow as distributed by the NOIRLab Astro Data Archive. NOIRLab is managed by the Association of Universities for Research in Astronomy (AURA) under a cooperative agreement with the National Science Foundation.

Based on observations obtained at the Southern Astrophysical Research (SOAR) telescope, which is a joint project of the Minist\'{e}rio da Ci\^{e}ncia, Tecnologia, Inova\c{c}\~{o}es e Comunica\c{c}\~{o}es (MCTIC) do Brasil, the U.S. National Optical Astronomy Observatory (NOAO), the University of North Carolina at Chapel Hill (UNC), and Michigan State University (MSU).

This paper includes data collected with the TESS mission, obtained from the MAST data archive at the Space Telescope Science Institute (STScI). Funding for the TESS mission is provided by the NASA Explorer Program. STScI is operated by the Association of Universities for Research in Astronomy, Inc., under NASA contract NAS 5–26555.  

This research made use of Lightkurve, a Python package for Kepler and TESS data analysis \citep{lightkurve}.  This research made use of Photutils, an Astropy package for detection and photometry of astronomical sources \citep{Photutils}.

%





\bibliography{references}{}
\bibliographystyle{aasjournal}




\newpage
\appendix

\begin{figure*}[h]
    \centering
    \includegraphics[width=\textwidth]{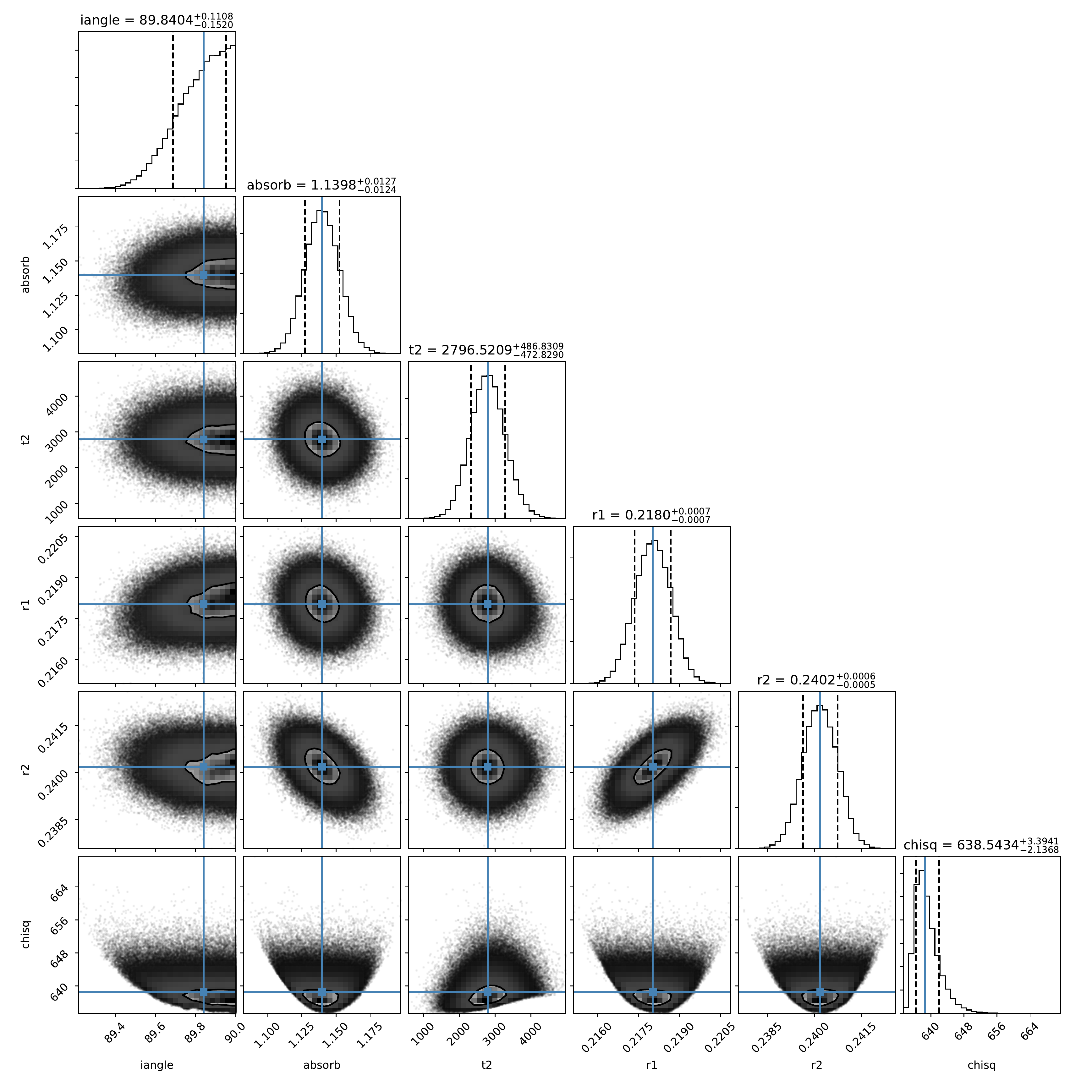}
    \caption{Corner plot of the most probable light curve solution for the SOAR/Goodman -- $B$ data.}
    \label{fig:mcmc_B}
\end{figure*}

\begin{figure*}[h]
    \centering
    \includegraphics[width=\textwidth]{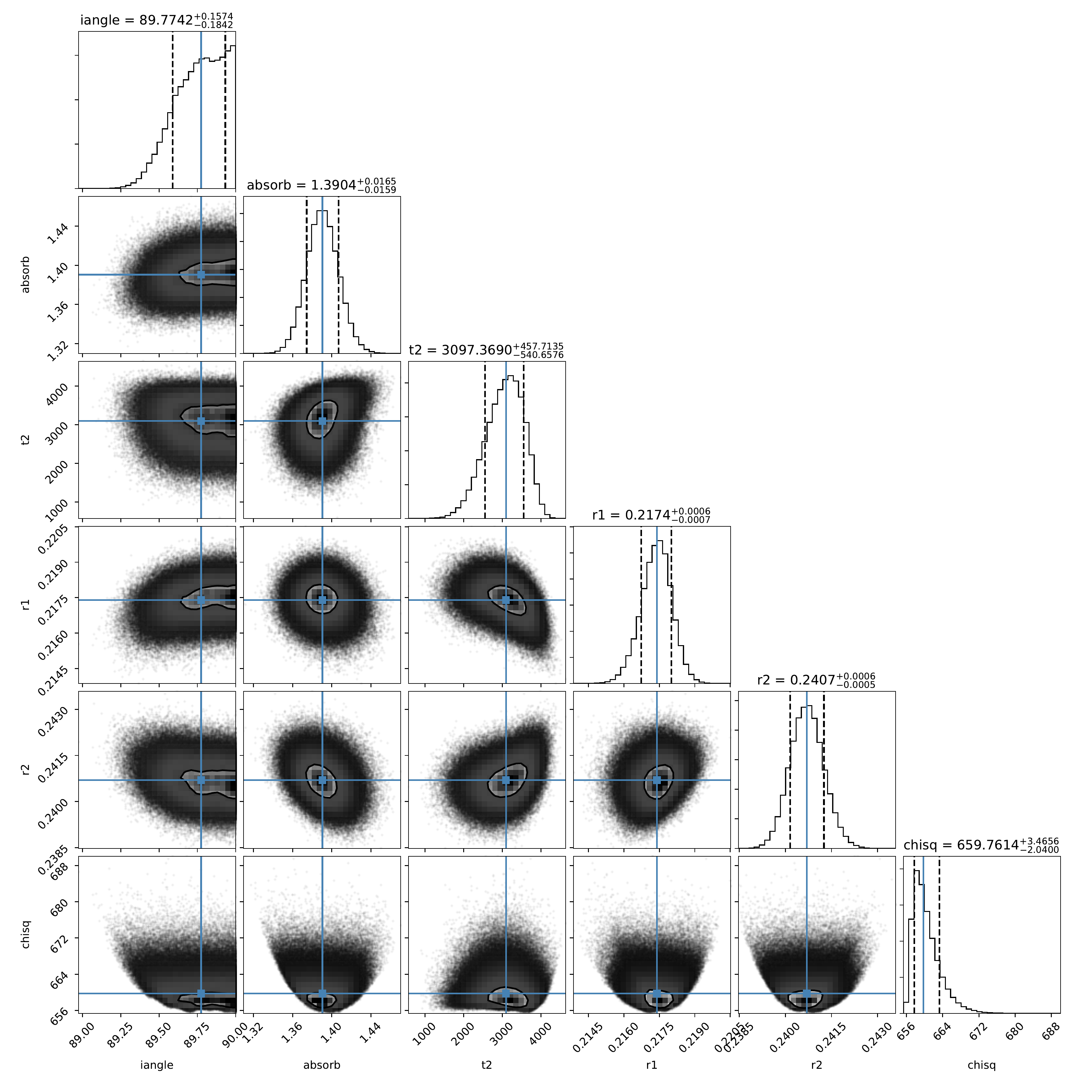}
    \caption{Corner plot of the most probable light curve solution for the SOAR/Goodman -- $R$ data.}
    \label{fig:mcmc_R}
\end{figure*}

Shown in Figures \ref{fig:mcmc_B} and \ref{fig:mcmc_R} are the corner plots for the SOAR/Goodman $B$ and $R$ light curve solutions, respectively, using the Python package \texttt{corner} \citep{Foreman-Mackey2016} for visualization.  Also, we give the full set of possible solutions from the light curve modeling in Table \ref{tab:q_solutions}.

\begin{table*}
    \centering
    \caption{All of the possible light curve solutions output by \texttt{LCURVE}.}
    
    \begin{tabular}{@{} lrrrrrr @{}}
    \toprule
    
    \multicolumn{1}{c}{\footnotesize{$q$}} & \multicolumn{1}{c}{\footnotesize{$a$}} &
    \multicolumn{1}{c}{\footnotesize{$M_{\rm sdB}$}} & \multicolumn{1}{c}{\footnotesize{$M_{\rm dM}$}} & \multicolumn{1}{c}{\footnotesize{$R_{\rm sdB}$}} & \multicolumn{1}{c}{\footnotesize{$R_{\rm dM}$}} &
    \multicolumn{1}{c}{\footnotesize{$\log(g)$}} \\
    
    \multicolumn{1}{c}{\scriptsize{}} & 
    \multicolumn{1}{c}{\scriptsize{[$R_{\odot}$]}} &
    \multicolumn{1}{c}{\scriptsize{[$M_{\odot}$]}} & 
    \multicolumn{1}{c}{\scriptsize{[$M_{\odot}$]}} & 
    \multicolumn{1}{c}{\scriptsize{[$R_{\odot}$]}} & 
    \multicolumn{1}{c}{\scriptsize{[$R_{\odot}$]}} &
    \multicolumn{1}{c}{\scriptsize{}} \\
    
    \midrule
    
    0.30 & $1.090\pm0.022$ & $0.826\pm0.050$ & $0.248\pm0.015$ & $0.234\pm0.005$ & $0.268\pm0.005$ & $5.616\pm0.009$  \\
    0.31 & $1.063\pm0.021$ & $0.760\pm0.046$ & $0.236\pm0.014$ & $0.231\pm0.005$ & $0.261\pm0.005$ & $5.591\pm0.009$  \\
    0.32 & $1.038\pm0.021$ & $0.701\pm0.042$ & $0.224\pm0.013$ & $0.224\pm0.004$ & $0.253\pm0.005$ & $5.585\pm0.009$  \\
    0.33 & $1.014\pm0.020$ & $0.649\pm0.039$ & $0.214\pm0.013$ & $0.219\pm0.004$ & $0.246\pm0.005$ & $5.571\pm0.009$  \\
    0.34 & $0.992\pm0.020$ & $0.603\pm0.036$ & $0.205\pm0.012$ & $0.214\pm0.004$ & $0.240\pm0.005$ & $5.558\pm0.009$  \\
    0.35 & $0.971\pm0.019$ & $0.561\pm0.034$ & $0.196\pm0.012$ & $0.209\pm0.004$ & $0.234\pm0.005$ & $5.547\pm0.009$  \\
    0.36 & $0.951\pm0.019$ & $0.523\pm0.031$ & $0.188\pm0.011$ & $0.205\pm0.004$ & $0.229\pm0.005$ & $5.531\pm0.009$  \\
    0.37 & $0.932\pm0.019$ & $0.489\pm0.029$ & $0.181\pm0.011$ & $0.201\pm0.004$ & $0.224\pm0.004$ & $5.521\pm0.009$  \\
    0.38$^{\dagger}$ & $0.914\pm0.018$ & $0.458\pm0.027$ & $0.174\pm0.010$ & $0.197\pm0.004$ & $0.219\pm0.004$ & $5.509\pm0.009$  \\
    0.39 & $0.897\pm0.018$ & $0.430\pm0.026$ & $0.168\pm0.010$ & $0.194\pm0.004$ & $0.215\pm0.004$ & $5.497\pm0.009$  \\
    0.40 & $0.881\pm0.018$ & $0.404\pm0.024$ & $0.162\pm0.010$ & $0.190\pm0.004$ & $0.210\pm0.004$ & $5.486\pm0.009$  \\
    0.41 & $0.865\pm0.017$ & $0.380\pm0.023$ & $0.156\pm0.009$ & $0.187\pm0.004$ & $0.206\pm0.004$ & $5.473\pm0.009$  \\
    0.42 & $0.851\pm0.017$ & $0.359\pm0.022$ & $0.151\pm0.009$ & $0.184\pm0.004$ & $0.203\pm0.004$ & $5.464\pm0.009$  \\
    0.43 & $0.837\pm0.017$ & $0.339\pm0.020$ & $0.146\pm0.009$ & $0.181\pm0.004$ & $0.199\pm0.004$ & $5.453\pm0.009$  \\
    0.44 & $0.824\pm0.016$ & $0.321\pm0.019$ & $0.141\pm0.008$ & $0.178\pm0.004$ & $0.195\pm0.004$ & $5.443\pm0.009$  \\
    0.45 & $0.811\pm0.016$ & $0.304\pm0.018$ & $0.137\pm0.008$ & $0.175\pm0.004$ & $0.192\pm0.004$ & $5.433\pm0.009$  \\
    
    
    
    \bottomrule
    
    \multicolumn{7}{@{}l}{\textbf{\small{Notes}}}\\
    \multicolumn{7}{@{}l}{\footnotesize{$^{\dagger}$ Most probable solution as outlined in the text}} 
    
    \end{tabular}
    
    \label{tab:q_solutions}
\end{table*}

\end{document}